\documentclass[journal,twoside]{IEEEtran}

\usepackage{cite}
\usepackage{amsmath,amssymb,amsfonts}
\usepackage{algorithmic}
\usepackage{graphicx}
\usepackage{textcomp}
\usepackage{multirow}
\usepackage{tabularx}
\usepackage{float}
\usepackage{stfloats}
\usepackage{ifpdf}
\usepackage{url}
\usepackage[OT1]{fontenc} 
\usepackage{hhline}
\usepackage{array}

\begin{document}
	\title{Characterization and Modeling of 0.18$\mu$m CMOS Technology at sub-Kelvin Temperature}
		
	\author{Tengteng Lu, Zhen Li, Chao Luo,
		Jun Xu, Weicheng Kong, and~Guoping~Guo,~\IEEEmembership{Member,~IEEE}
		\thanks{Manuscript received XX XX, 2008; revised XX XX, 2018. This work was supported in part by the National Key Research and Development Program of China (Grant No. 2016YFA0301700), in part by the National Natural Science Foundation of China (Grants No. 11625419), in part by the Anhui initiative in Quantum information Technologies (Grants No. AHY080000) and this work was partially carried out at the USTC Center for Micro and Nanoscale Research and Fabrication. (corresponding author: Guoping Guo)}%
		\thanks{T Lu, Z Li, C Luo are with the Key Laboratory of Quantum Information, University of Science and Technology of China, Hefei, Anhui 230026, China, and also with the Department of Physics, University of Science and Technology of China, Hefei, Anhui 230026, China.}
		\thanks{J Xu is with the Department of Physics, University of Science and Technology of China, Hefei, Anhui 230026, China.}
		\thanks{W Kong is with the Origin Quantum Computing Company Limited, Hefei, Anhui 230026, China.}%
		\thanks{G Guo is with the Key Laboratory of Quantum Information, University of Science and Technology of China, Hefei, Anhui 230026, China(e-mail: gpguo@ustc.edu.cn).}}%
		
	\markboth{Journal of the Electron Devices Society,~Vol.~XX, No.~XX, XXXX~2018}
	{Tengteng Lu  \MakeLowercase{\textit{et al.}}: Characterization and Modeling of 0.18$\mu$m CMOS Technology at Sub-kelvin Temperature}	
	\maketitle
	
\begin{abstract}

Previous cryogenic electronics studies are most above 4.2K. In this paper we present the cryogenic characterization of a 0.18$\mu$m standard bulk CMOS technology(1.8V and 5V) at sub-kelvin temperature around 270mK. PMOS and NMOS devices with different width to length ratios(W/L) are tested and characterized under various bias conditions at temperatures from 300K to 270mK. It is shown that the 0.18$\mu$m standard bulk CMOS technology is still working at sub-kelvin temperature. The kink effect and current overshoot phenomenon are observed at sub-kelvin temperature. Especially, current overshoot phenomenon in PMOS devices at sub-kelvin temperature is shown for the first time. The transfer characteristics of large and thin-oxide devices at sub-kelvin temperature are modeled using the simplified EKV model. This work facilitates the CMOS circuits design and the integration of CMOS circuits with silicon-based quantum chips at extremely low temperatures. 
\end{abstract}
		
\begin{IEEEkeywords}
	Cryogenic CMOS, characterization, modeling, kink effect, current overshoot, sub-kelvin temperature.
\end{IEEEkeywords}

	%
	\IEEEpeerreviewmaketitle

\section{Introduction}
\IEEEPARstart{C}{ryogenic} CMOS(cryo-CMOS) has been researched in recent years due to the rise of quantum chip and quantum computer research\cite{IEEEexample:1}\cite{Sebastiano:2017:CEC:3061639.3072948}. CMOS circuits integrated with quantum bits on the same substrate or standalone CMOS circuits working at extremely low temperatures can improve the scalability of quantum chips, system integration and performance\cite{7993523}\cite{7838410}. However, cryo-CMOS circuits face several changes such as refrigeration power, interconnection and device modeling.

The applicable temperature for BSIM3 or BSIM4 model is from - 40℃ to 125 ℃ in general. The characteristics of MOSFET devices change due to freeze-out effect at low temperatures. At present, CMOS technologies ranging from 0.35$\mu$m to 28nm have been characterized and modeled at the liquid nitrogen temperature(77K) and the liquid helium temperature(4.2K)\cite{8370029,1742-6596-834-1-012002,MARTIN2011115}. However, the characteristic testing and modeling at the sub-kelvin temperature are lacking\cite{1742-6596-834-1-012005}.

The paper contains the characteristics of 0.18$\mu$m MOSFET devices, kink effect and current overshoot phenomenon of MOSFET at sub-kelvin temperature. The EKV model for transfer characteristics of large size devices is provided, which can be used for estimating power consumption of devices and the simulation of CMOS circuits design at sub-kelvin temperature.

\section{Cryogenic Measurement Setup}
Measurements of CMOS transistors were performed with two different gate oxide thicknesses(TOX) and a wide range of device sizes at around 270mK, as shown in Table \ref{table1}. The tested chip and chip-carrier were connected with Al-wire bonding. The $^{3}$He refrigerator was used to cool devices. The devices under test(DUT) were put into the dual in-line package(DIP) socket at the bottom of the refrigerator and connected to BNC connectors at the top of the refrigerator via cables. An internal vacuum chamber was built around the sample, and the sample was inserted into the cryostat to cool down to around 270mK.All MOSFET electrical measurements were performed using the Keysight B1500A semiconductor device analyzer. The interconnection between the DUT and B1500A was realized through BNC cables and BNC Triax to BNC adapters.
\begin{table}[htb]
	\centering
	\renewcommand{\arraystretch}{1.3}
	\caption{SUMMARY OF CHARACTERIZED DEVICES}
	\begin{tabular}{|c|c|c|c|c|}
		\hline
		Technology& \multicolumn{4}{c|}{SMIC 0.18$\mu$m Bulk CMOS 1P6M Process}\\
		\hline
		Voltage and TOX& \multicolumn{2}{c|}{1.8V Thin}& \multicolumn{2}{c|}{5V Thick}\\
		\hline
		Type& \multicolumn{1}{c|}{NMOS}& \multicolumn{1}{c|}{PMOS}& \multicolumn{1}{c|}{NMOS}& \multicolumn{1}{c|}{PMOS}\\
		\hline
		\multirow{5}{*}{W/L[$\mu$m/ $\mu$m]}
		~& 10/10& 10/10 & 10/10& 10/10\\
		\cline{2-5}
		~& 10/0.6& 10/0.6& 10/2& 10/2\\
		\cline{2-5}
		~& 10/0.2& 10/0.18& 10/0.65& 10/0.5\\
		\cline{2-5}
		~& 10/0.18& 10/0.16& 10/0.5& 10/0.45\\
		\cline{2-5}
		~& 10/0.16&  & 0.3/0.6&  \\
		\hline
	\end{tabular}
	\label{table1}
\end{table}
	
Testing of the characteristics at around 270mK should limit the power of devices, limited by the cooling power. For 1.8V MOSFET, transfer characteristics in linear$(|V_{DS}|=50mV )$  and saturation regions $(|V_{DS}|=1.8V )$ for different substrate bias voltages were measured on various devices in Table \ref{table1}; output characteristics in zero substrate bias $(|V_{BS}|=0V )$ and reverse bias voltage $(|V_{BS}|=1.8V )$ for different gate voltages were measured. For 5V MOSFET, bias voltages changed. Characteristics in other bias voltages were also measured at around 270mK. In addition to the above measurements, the output characteristics were measured which were in different hold-time and delay time. 
\section{Characterization}
\subsection{Transfer Characteristics}
Fig.1 shows the transfer characteristics of large and small thin TOX MOSFET devices. Fig.2 shows the transfer characteristics of the thick TOX NMOS (W/L=0.3$\mu$m/0.6$\mu$m) and PMOS (W/L =10$\mu$m/10$\mu$m) in liner region. The drain current is still depending on the $V_{GS}$  and W/L. The threshold voltage $(V_{th})$ changes with substrate bias voltage. Transfer characteristics are similar to the usual results for MOSFET. Fig. 1(a,b) give transfer characteristics of the large NMOS device at sub-kelvin temperature. For the small NMOS device in Fig.1(c,d), the $V_{th}$ decreases rapidly because of drain induced barrier lowering(DIBL) effect\cite{845589}\cite{QU2011298}. The DIBL effect is also shown in Fig.1(g,h).

\begin{figure*}[!t]
	\centering
	\includegraphics[width=516pt]{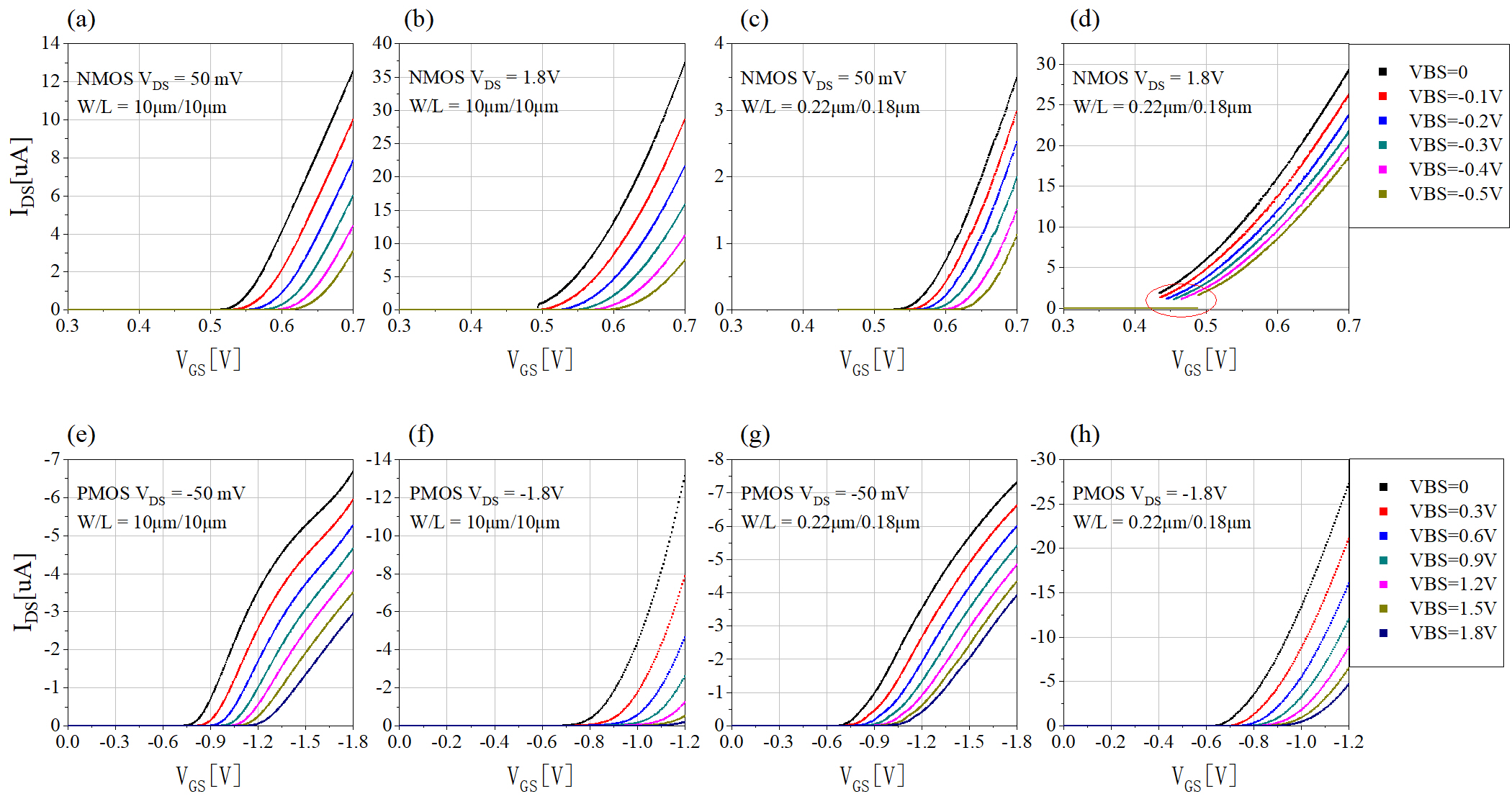}
	\caption{Transfer characteristics of thin TOX MOSFET devices at sub-kelvin temperature. (a-d) Transfer characteristics of the large size NMOS device(W/L = 10$\mu$m/10$\mu$m) and the small size NMOS device(W/L = 0.22$\mu$m/0.18$\mu$m)，V$_{GS}$=0.3$V$$\rightarrow$0.7$V$ step=1$mV$, V$_{BS}$=0$V$$\rightarrow$-0.5$V$ step=-0.1$V$, V$_{DS}$=50$mV$ or 1.8$V$. (e-h) Transfer characteristics of the large size PMOS device(W/L = 10$\mu$m/10$\mu$m) and the small size PMOS device(W/L = 0.22$\mu$m/0.18$\mu$m)，V$_{GS}$=0$V$$\rightarrow$-1.2$V$ or -1.8$V$ step=-5$mV$,  V$_{BS}$=0$V$$\rightarrow$1.8$V$ step=0.3$V$, V$_{DS}$=-50$mV$ or -1.8$V$.}
	\label{fig1}
\end{figure*}
\begin{figure}[!t]
	\centering
	\includegraphics[width=1\columnwidth]{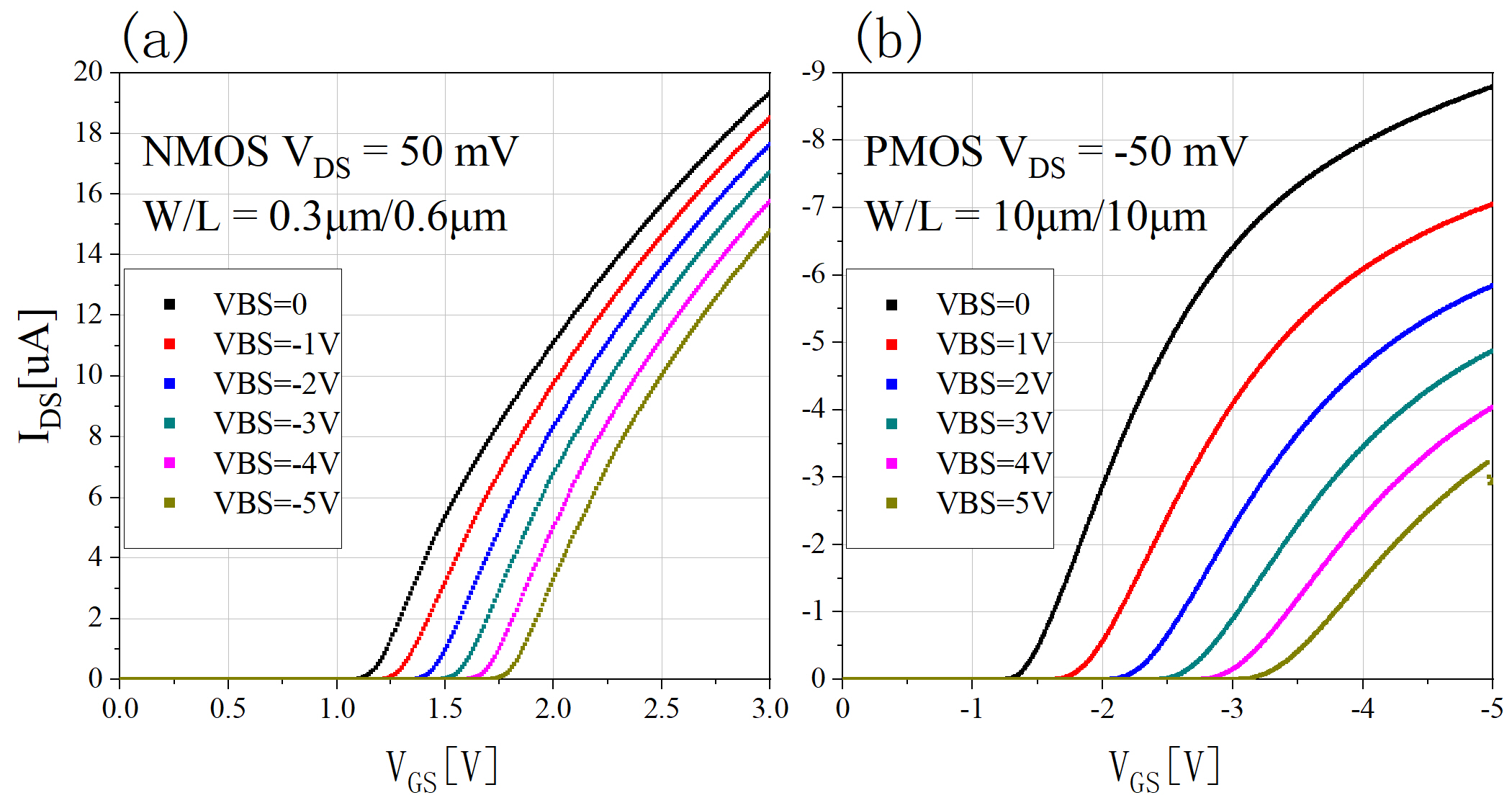}
	\caption{Transfer characteristics of thick TOX MOS at sub-kelvin temperature. (a) Transfer characteristics of the small size NMOS device(W/L = 0.3$\mu$m/0.6$\mu$m), V$_{GS}$=0$V$$\rightarrow$3$V$ step=10$mV$, V$_{BS}$=0$V$$\rightarrow$-5$V$ step=-1$V$, V$_{DS}$=50$mV$; (b) Transfer characteristics of the large size PMOS device(W/L = 10$\mu$m/1$\mu$μm), V$_{GS}$=0$V$$\rightarrow$-5$V$ step=-10$mV$, V$_{BS}$=0$V$$\rightarrow$5$V$ step=1$V$, V$_{DS}$=-50$mV$.}
	\label{fig2}
\end{figure}

An abnormal phenomenon about current step is shown in Fig.1(d); for $V_{BS}=0$ in Fig.1(b), we can observe current step phenomenon also. When the gate to source voltage $(V_{GS})$ reaches $V_{th}$, $I_{DS}$ suddenly changes into microamperes($\mu$A) current. The abnormal phenomenon prohibits the use of small NMOS devices in saturation region for extremely low-power circuits. 
	
\subsection{Output Characteristics and Kink Effect}
We have plotted in Fig.3(a-d) output characteristics of thin and thick TOX MOSFET devices, obtained at sub-kelvin temperature. We choose the current-limiting method to keep the temperature stable. These curves (forward and backward scan) in Fig.3 show the kink effect. The kink effect can occur on MOSFET devices at sub-kelvin temperatures when the $|V_{DS}|$ becomes sufficiently high\cite{BALESTRA1987321}. For bulk silicon NMOS devices at low temperatures, channel electron can gain enough energy and produce electron-hole pairs by impact ionization when the $V_{DS}$ is higher than a certain value. The electrons pass through the channel to the drain, and the holes migrate towards the floating substrate. The floating substrate potential increases due to the accumulation of holes, until the substrate-source junction forms forward bias\cite{BALESTRA1987321}\cite{2052}. The $V_{th}$ reduces with the increase of substrate potential. As the $|V_{DS}|$ increases, the $V_{th}$ decreases, resulting in the increase of drain-source current $(I_{DS})$.

\begin{figure*}[!t]
	\centering
	\includegraphics[width=516pt]{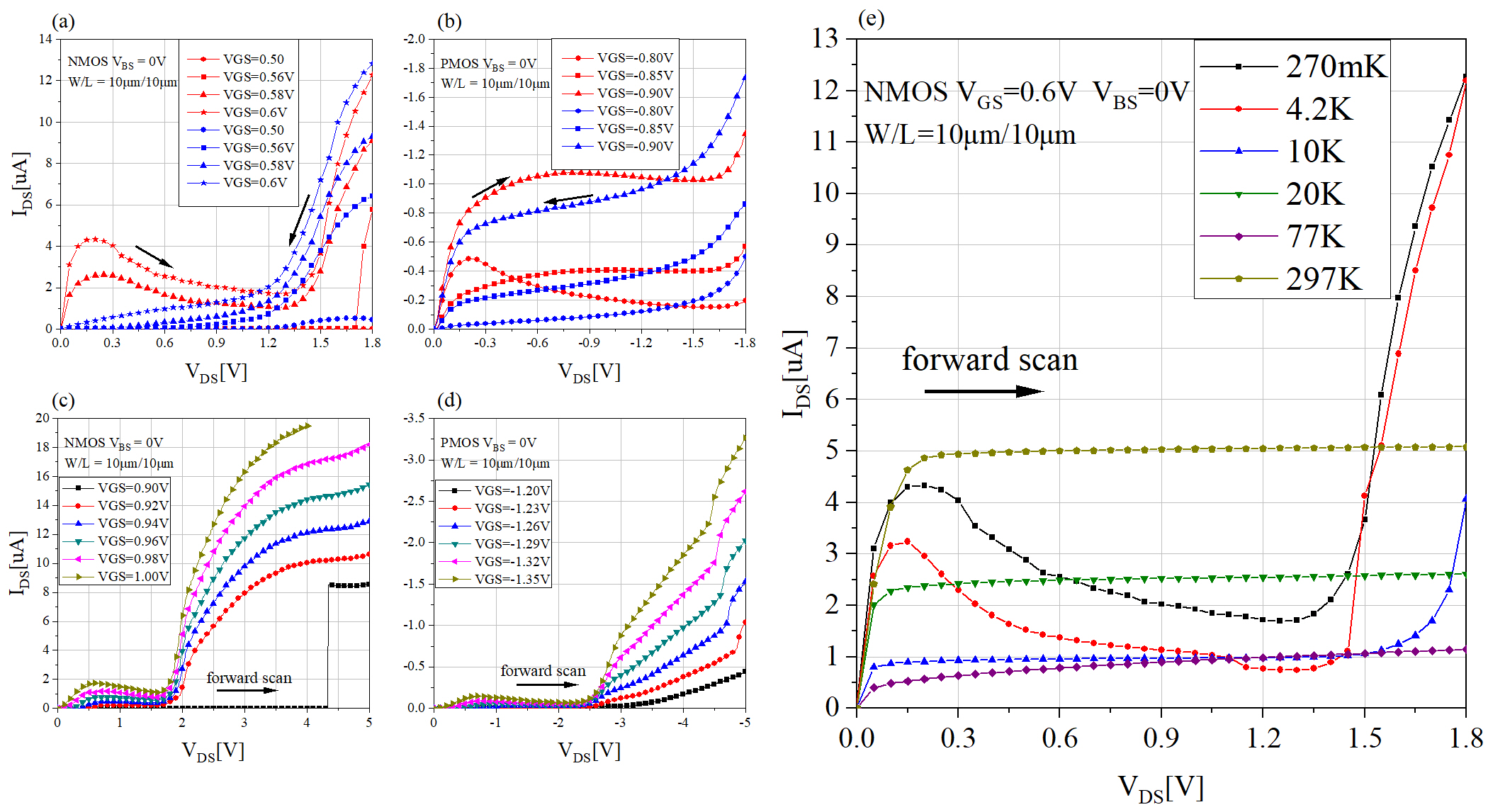}
	\caption{(a,b)Output characteristics of thin TOX MOS at sub-kelvin temperature, the forward(backward) scan data is represented by red(blue) lines and dots. (c,d)Output characteristics of thick TOX MOS at sub-kelvin temperature with forward scan. (e)Output characteristics of the large thin TOX NMOS device(W/L = 10$\mu$m/10$\mu$m) at temperatures from 297K down to 270mK, V$_{DS}$=0$V$$\rightarrow$1.8$V$ step=50$mV$, V$_{GS}$=0.6$V$,V$_{BS}$=0$V$, delay time=0$s$.}
	\label{fig3}
\end{figure*}
	
Compared to PMOS devices, the kink effect in NMOS devices appears at lower $|V_{DS}|$ voltage, and the amplification of NMOS is bigger. Considering to the higher impact-ionization multiplication factor of electrons compared to holes, NMOS devices are more affected at sub-kelvin temperature compared to PMOS devices\cite{5109798}.
	
Besides, we observed the current overshoot phenomenon in Fig.3 when the scan model is forward scan. For forward scan results of thin TOX PMOS, the current overshoot occurs obviously while the test begins. The kink effect and current overshoot phenomenon at cryogenic temperatures are related to varying $V_{th}$ with $V_{DS}$. As shown in Fig.3(e), the kink effect and current overshoot phenomenon attenuate and disappear with the temperature increasing. 
	
\subsection{Current Overshoot and Discussion}
The current overshoot phenomenon occurred in Fig.3(e) at temperatures below 10K other than 15K\cite{LYSENKO2000735}. This phenomenon is related to the charge state of traps in the Si/SiO$_{2}$ interface. The Si/SiO$_{2}$ interface traps are silicon dangling-bonds at the interface. In the case of on state NMOS devices (forward scan), it can be seen that positive charging of traps leading to the noticeable increase of the current at the beginning of curves in Fig.4(a). The $V_{th}$ decreases in the liner region for NMOS devices owing to positive charge trapped in the interface traps. When $|V_{DS}|$ is sufficiently high after reaching the pinch-off point, the trapped charges in the interface can be released at sub-kelvin temperature\cite{LYSENKO2000735}. The $V_{th}$ increases and the amplitude of $I_{DS}$ curves decreases in the saturation region as the $|V_{DS}|$ increases, as shown in Fig.4(a,b).

\begin{figure*}[!t]
	\centering
	\includegraphics[width=516pt]{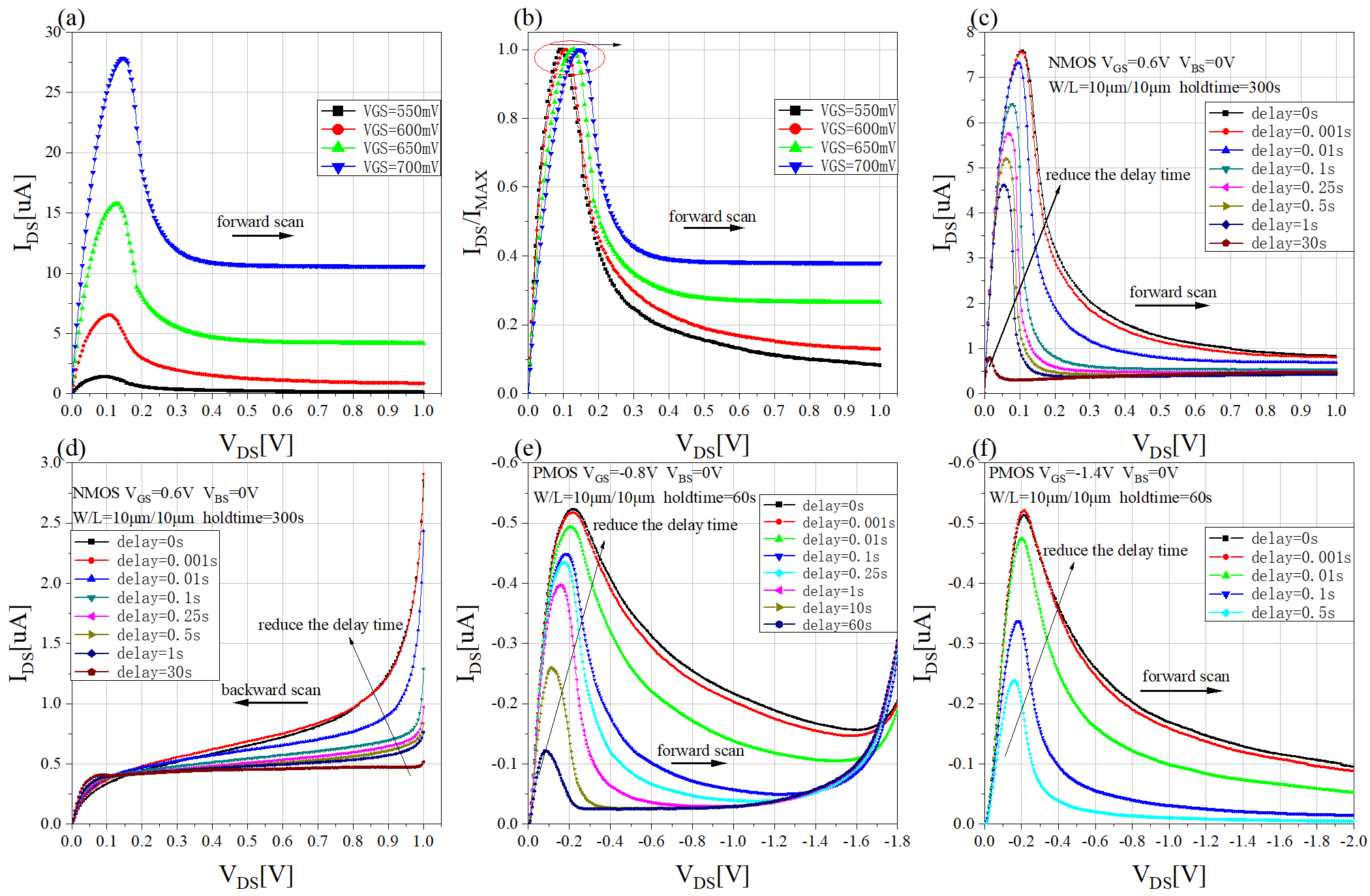}
	\caption{ Current overshoot phenomenon in the thin TOX NMOS(a-d) device(W/L = 10$\mu$m/10$\mu$m) and PMOS(e,f) devices(W/L = 10$\mu$m/10$\mu$m) and at sub-kelvin temperature. (a) I$_{DS}$-V$_{DS}$ curves for different V$_{GS}$; (b) I$_{DS}$/I$_{MAX}$-V$_{DS}$ curves, normalized by I$_{MAX}$ which is the maximum current for each curves in (a); (c) forward scan with different delay time; (d) backward scan with different delay time; (e) forward scan for the large thin TOX PMOS device with different delay time; (f) forward scan for the large thick TOX PMOS device with different delay time.}
	\label{fig4}
\end{figure*}

We measured the current overshoot with various delay time and two scan models. In Fig.4(c), the maximum amplitude of curves is 7.59$\mu$A for 0$s$ delay time, and the peak value decreases when the scan delay time increases. Predictably, the peak will disappear on the sufficient scan delay time. We do not observe the current overshoot in Fig.4(d), plotted results with backward scan model. For PMOS devices, the current overshoot phenomenon also occurred at sub-kelvin temperatures in Fig.4(e,f). The relation between the current overshoot phenomenon and the scan delay time implies the charge state of traps in the Si/SiO$_{2}$ interface\cite{LYSENKO2000735}\cite{1742-6596-834-1-012005}. 
	
\section{Modeling}
\subsection{EKV Model for Large Devices}
The EKV MOSFET model is applicable to MOSFET devices in low-voltage and low-current applications\cite{Enz1995}. And this model is compact and accurate with a few parameters\cite{1314621}\cite{Angelov06mosfetsimulation}. The output characteristics of MOFET devices are abnormal at sub-kelvin temperature because of kink effect and current overshoot, and the data of thick TOX large devices is insufficient by the limitation of the cooling power. Given this, we choose the EKV 2.6 model to describe transfer characteristics of thin TOX large MOSFET devices. For large MOSFET devices, the intrinsic model parameters in Table \ref{table2} can describe transfer characteristics accurately, and the short and narrow channel effects are ignored. The value of parameters in Table \ref{table2} is for thin TOX PMOS devices.
\begin{table}[!hbp]
	\centering
	\renewcommand{\arraystretch}{1.3}
	\caption{THE INTRINSIC MODEL PARAMETERS}
	\begin{tabular}{ccc}
		\hline
		\hline
		Name(unit)&Description&Value\\
		\hline
		TOX($m$)& oxide thickness & 4.27e-009\\ 
		XJ($m$) & junction depth & 1.6e-007\\
		DW($m$) & channel width correction & -1.6e-008\\
		DL($m$) & channel length correction & 7.6e-008\\
		NSUB($cm^{-3}$) & channel doping & 1.0196e+017\\
		UCRIT($V/m$) & longitudinal critical field & 2.0e+6\\
		VTO($V$) & nominal threshold voltage & - \\
		KP($A/V^{2}$) & transconductance parameter & - \\
		GAMMA($\sqrt{V}$) & body effect parameter & - \\
		THETA($1/V$) & mobility reduction coefficient & - \\
		\hline
		\hline
	\end{tabular}
	\label{table2}
\end{table}		
	
Parameters of TOX, XJ, DW, DL, and NSUB are from BSIM4 model\cite{ref18}, and the parameter(UCRIT) can be set as default. The value of PHI(bulk Fermi potential) parameter is calculated with NSUB parameter. The value of $T_{nom}$ is -272.88K(270mK), which is the nominal temperature of model parameters. We ignore the intrinsic parameters temperature dependence in EKV model because $T_{nom}$ is set as 270mK equal to test temperature. Equations of interest are presented below\cite{Angelov06mosfetsimulation}\cite{EKV}\cite{SPIE}.

\begin{equation}
	PHI = 2V_{t}\times \ln (\frac{NSUB \times 10^6}{n_{i}(T_{nom})})
\end{equation}
	where $n_{i}(T_{nom})$ is intrinsic carrier concentration,
	\begin{equation}
		V_{t} = \frac{kT}{q}
	\end{equation}
	The expression of $I_{DS}$ is 
	\begin{equation}
	I_{DS} = I_{F} - I_{R}
	\end{equation}
	$I_{F}$ is the forward current and $I_{R}$ is the reverse current.
	\begin{equation}
	I_{F(R)} = I_{S} \times \ln^2 [1+\exp(\frac{V_{G}-VTO-nV_{S(D)}}{2nV_{t}})] 
	\end{equation}
	where the specific current 
	\begin{equation}
	I_{S} = 2n\beta V_{t}, n = 1+\frac{GAMMA}{2\sqrt{V_{p}+PHI+4V_{t}}}
	\end{equation}
	Considering the mobility reduction due to vertical field,
	\begin{equation}
	\beta = KP\times \frac{L_{eff}}{W_{eff}}\times \frac{1}{1+THETA\times V_{P}}
	\end{equation}
	where $V_{P}$ is the pinch-off voltage for large devices.
	\begin{figure}[!h]
		\centering
		\includegraphics[width=1\columnwidth]{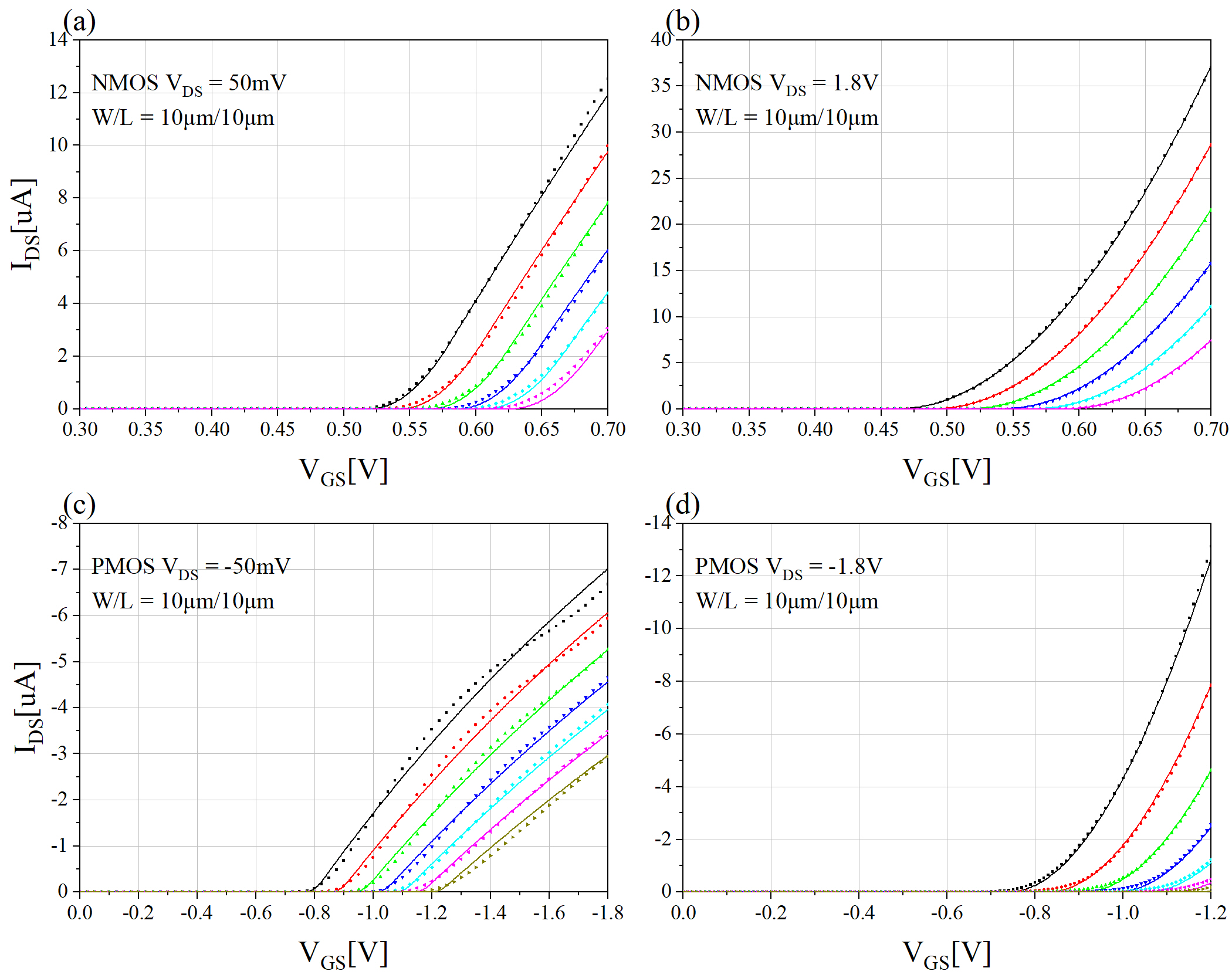}
		\caption{ I$_{DS}$-V$_{GS}$ curves of large thin TOX NMOS(a,b) and PMOS(c,d) devices at sub-kelvin temperature measured(symbols) and simulated(solid lines). The bias condition is same to \ref{fig1}. Model parameters are given in Table \ref{table3}(NMOS) and Table \ref{table4}(PMOS).}
		\label{fig5}
	\end{figure}	
	
\begin{table}[!h]
	\centering
	\renewcommand{\arraystretch}{1.3}
	\caption{PARAMETERS FOR THE LARGE THIN TOX NMOS DEVICE}
	\begin{tabular}{ccc}
		\hline
		Region & liner & saturation \\
		\hline
		VTO($V$) & 0.5208 & 0.46403 \\
		\hline
		GAMMA($\sqrt{V}$) & 0.50534 & 0.6024 \\
		\hline	
		KP($A/V^{2}$) & 0.0017217 & 0.0017958 \\
		\hline
		THETA($1/V$) & 0.462 & 0.31198 \\
		\hline
		RMS(\%) & 1.05 & 0.19\\
		\hline
	\end{tabular}	
	\label{table3}
\end{table}
\begin{table}[!h]
	\centering
	\renewcommand{\arraystretch}{1.3}
	\caption{PARAMETERS FOR THE LARGE THIN TOX PMOS DEVICE}
	\begin{tabular}{ccc}
		\hline
		Region & liner & saturation \\
		\hline
		VTO($V$) & -0.7676 & -0.74 \\
		\hline
		GAMMA($\sqrt{V}$) & 0.66736 & 0.675 \\
		\hline	
		KP($A/V^{2}$) & 0.00018393 & 0.0001806 \\
		\hline
		THETA($1/V$) & 0.32242 & 0.37118 \\
		\hline
		RMS(\%) & 1.52 & 0.38\\
		\hline
	\end{tabular}	
	\label{table4}
\end{table}

\subsection{Parameters Extraction and Modeling}
The parameters extraction is performed by BSIMProPlus with semi-empirical methods. We import the test data of thin TOX large MOEFET devices into the software and set value of TOX, XJ etc. The degree of agreement(described as RMS Error) between the simulation data and the test data changes with VTO, KP, GAMMA and THETA. After these parameters have been adjusted properly, the RMS Error declines to less than 5\% (Fig.5).
	
The value of $\vert$VTO$\vert$ in Table \ref{table3} and Table \ref{table4} decreases from liner region to saturation region, which means the $|V_{th}|$ of large devices varies with operating modes. The explanation of the kink effect is validated. Comparing the value of parameters in liner region with the value of parameters in saturation region, the numerical difference of NMOS is greater than PMOS. It is proved that NMOS devices are more affected at sub-kelvin temperature compared to PMOS devices. 

\section{Conclusion}
The SMIC 0.18$\mu$m bulk CMOS technology is tested and characterized at sub-kelvin temperature. It is demonstrated that MOSFET devices are still operating at sub-kelvin temperature. The kink effect and current overshoot phenomenon have been shown and explained. The EKV2.6 library with parameters extracted from transfer characteristics of thin TOX large devices at sub-kelvin temperature. The obtained results have proven the EKV2.6 library extracted is suited to describe transfer characteristics of thin TOX large devices at 270mK. This work contributes to research on characteristics of devices and cryogenic CMOS circuits design for quantum chips. 
	
\section*{Acknowledgment}
	
The authors would like to thank SMIC for devices fabrication and software support. 
		
\ifCLASSOPTIONcaptionsoff
\newpage
\fi

	
	
\bibliographystyle{IEEEtran}
\bibliography{IEEEabrv,IEEEexample}

%
	
\begin{IEEEbiography}[{\includegraphics[width=1in,height=1.25in,clip,keepaspectratio]{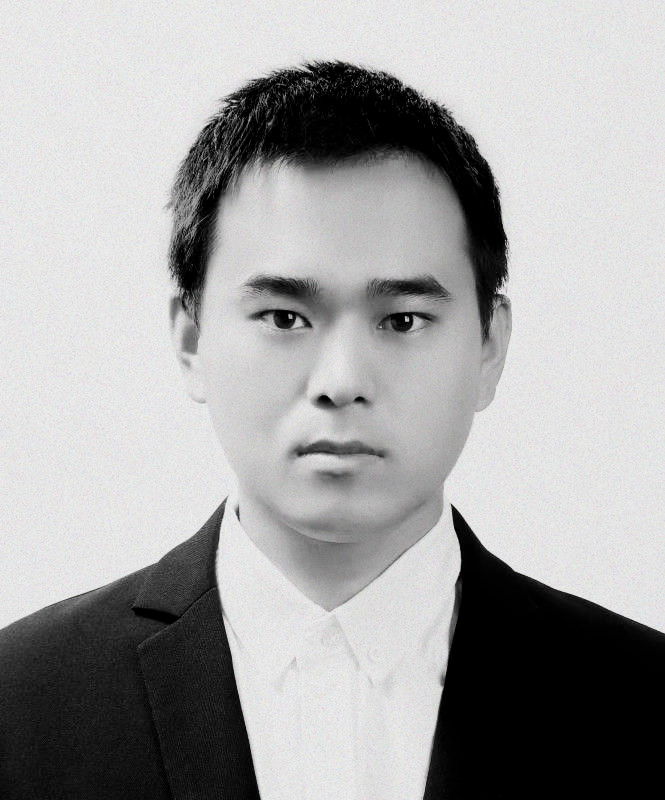}}]
{Teng-Teng Lu} recived the B.S. degree in physics from the University of Science and Technology of China, Hefei, China, in 2015. He is currently pursuing the Ph.D. degree in microelectronics and solid state electronics from University of Science and Technology of China, Hefei, China.
		
His current research interests include modeling of CMOS technologies and analog design at cryogenic temperatures.
\end{IEEEbiography}
	
\begin{IEEEbiography}[{\includegraphics[width=1in,height=1.25in,clip,keepaspectratio]{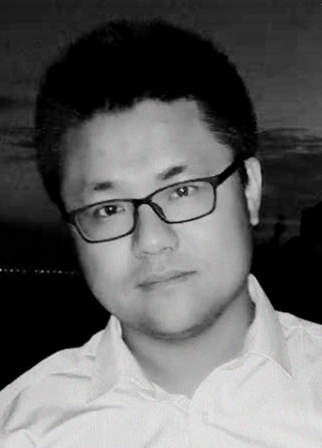}}]
{Zhen Li} received the B.S. degree in physics from University of Science and Technology of China(USTC), Hefei, China in 2014. He is currently pursuing the Ph.D. degree in microelectronics and solid state electronics from University of Science and Technology of China, Hefei, China.
		
He is currently a Research Assistant with the Key Laboratory of Quantum Information, CAS, USTC, Hefei, China since 2016. His research interests include cryogenic electronics and quantum computing.
\end{IEEEbiography}
	
\begin{IEEEbiography}[{\includegraphics[width=1in,height=1.25in,clip,keepaspectratio]{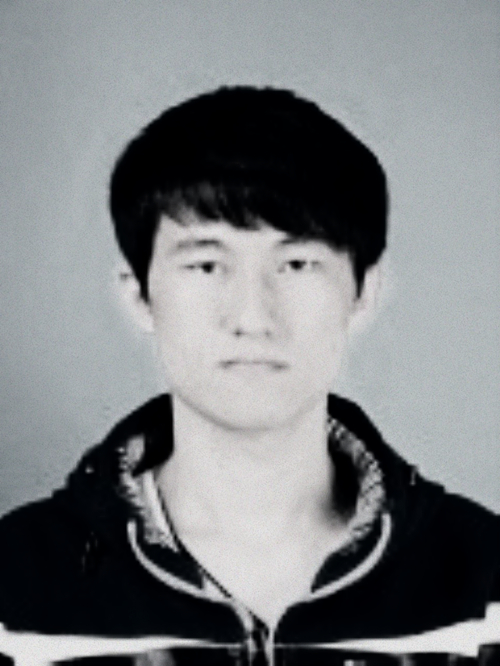}}]
{Chao Luo} received the B.S. degrees in physics from the University of Science and Technology of China, Hefei, China in 2014. He is currently pursuing the Ph.D. degree in microelectronics and solid state electronics from University of Science and Technology of China, Hefei, China.
		
His research interests include cryogenic electronics and the electronic interface operating at cryogenic temperature for quantum computing.
\end{IEEEbiography}
	
\begin{IEEEbiography}[{\includegraphics[width=1in,height=1.25in,clip,keepaspectratio]{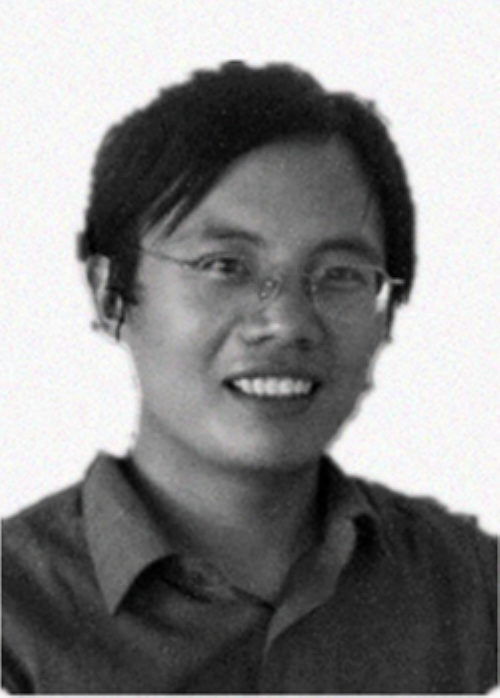}}]
{Jun Xu} received the B.S., M.S and Ph.D degrees in physics fron University of Science and Technology of China, Hefei, China in 2000, 2002 and 2006, respectively.
		
He is a lectorate in department of physics in University of Science and Technology of China, Hefei, China. His research interests include photoelectric devices and sensing measurement.
		
\end{IEEEbiography}
	
\begin{IEEEbiography}[{\includegraphics[width=1in,height=1.25in,clip,keepaspectratio]{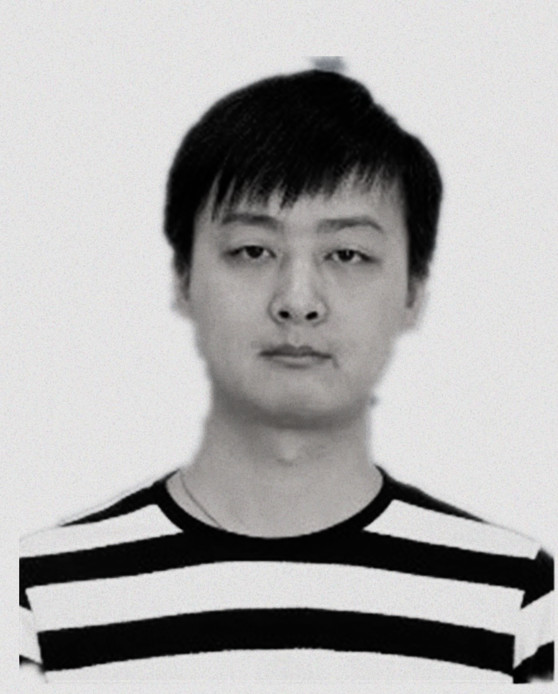}}]
{Weicheng Kong} received B.S. and Ph.D. degree in physics from University of Science and Technology of China, Hefei, China in 2013 and 2018, respectively.
		
He is currently the executive director of Origin Quantum Computing Co. Ltd, Hefei, China. His research interests include quantum manipulation and measurements of mesoscopic and microscopic quantum devices.
		
\end{IEEEbiography}
	
\begin{IEEEbiography}[{\includegraphics[width=1in,height=1.25in,clip,keepaspectratio]{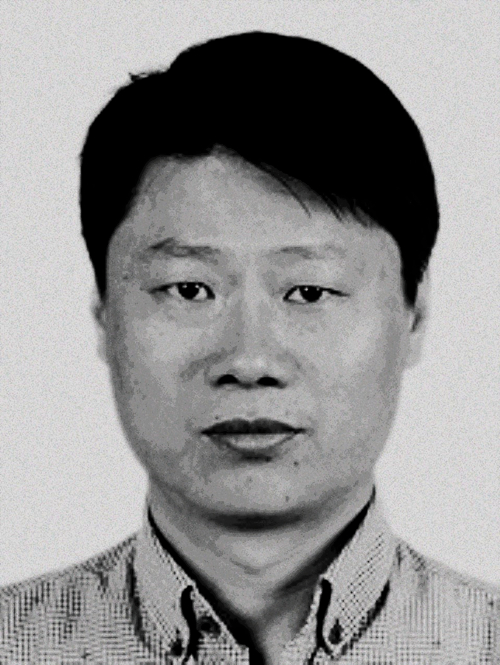}}]
{Guoping Guo} received the B.S. and Ph.D. degrees in physics fron University of Science and Technology of China, Hefei, China in 2000 and 2005, respectively.
		
He is currently a professor of physics and the Associated Director of the Key Laboratory of Quantum Information, CAS, University of Science and Technology of China, Hefei, China. His research interests include quantum computing and cryogenic electronics.
		
\end{IEEEbiography}
	
	\vfill
	

\end{document}